\documentclass[twocolumn,aps]{revtex4}
\usepackage{graphicx}

\def\be{\begin{equation}}
\def\ee{\end{equation}}
\def\bea{\begin{eqnarray}}
\def\eea{\end{eqnarray}}

\def\C{{\cal C}}

\begin{document}

\title{Optimality of minimum-error discrimination by the no-signalling condition}

\author{Joonwoo Bae$^{1}$, Jae-Weon Lee$^{1}$, Jaewan Kim$^{1}$ and Won-Young Hwang$^{2}$}

\affiliation{$^{1}$School of Computational Sciences, Korea
Institute for Advanced Study, Seoul 130-012, Korea\\
$^2$Department of Physics Education, Chonnam National University,
Kwangjoo 500-757, Korea}
\date{\today }

\begin{abstract}
In this work we relate the well-known no-go theorem that two
non-orthogonal (mixed) quantum states cannot be perfectly
discriminated, to the general principle in physics, the
no-signalling condition. In fact, the minimum error in
discrimination between two quantum states is derived from the
no-signalling condition.
\end{abstract}


\maketitle

\section{Introduction}

Two non-orthogonal quantum states cannot be discriminated with
certainty, while the discrimination error can be made smaller as
their copies are provided. This leads to one of the well-known
no-go theorems, that quantum states cannot be copied with
certainty \cite{no cloning}, although approximate quantum cloning
is possible with the use of quantum operations and ancillary
quantum systems \cite{buzek}. Interestingly, the impossibility of
perfect quantum cloning can be connected to the no-signalling
principle in physics, which dictates that information cannot be
sent faster than light. As a consequence, quantum communication
that makes use of (non-local) quantum correlations cannot be
performed faster than light.

In fact, the relation between the no-cloning theorem and the
no-signalling constraint has been established in both qualitative
and quantitative terms. For instance, it has been shown in Ref.
\cite{cons} that any no-signalling theory predicting non-locality,
i.e. violation of Bell inequalities, has a no-cloning theorem.
This implies that quantum theory necessarily has the no-cloning
theorem. In addition, the optimal fidelity of approximate quantum
cloning has been derived by applying the no-signalling condition
to the cloning process of quantum states \cite{nocloning-gisin},
provided that corresponding quantum operations are positive
\cite{bruss}.

On the other hand, the impossibility of perfect state
discrimination can also be connected to the no-signalling
condition in a qualitative way, at least through the existing
relation that the no-cloning theorem is implied by the
no-signalling constraint. Now the question naturally arises
whether there is a quantitative connection as well. In fact,
recent progress along this line has shown that the no-signalling
condition would imply the optimality of state discrimination
\footnote{Note that, when no-signalling constraint is considered
in state discrimination, the measurement postulate in quantum
theory such as positive-operator-valued-measure is assumed.},
constrained to those figures of merit such as minimum-error
discrimination \cite{hwang}, unambiguous state discrimination
\cite{steph1}, and maximum confidence measurement \cite{steph2}.
It is remarkable that the optimality of state discrimination in
different figures of merit can be derived from the single
condition, no-signalling constraint, despite the fact that quantum
theory is not a maximally non-local theory \cite{pr}.


In this work, we relate the no-signalling condition to
minimum-error state discrimination of ensembles of quantum states
i.e. mixed states. The novelty of this work is that two quantum
states to be discriminated between are not purely quantum but
mixtures of pure quantum states. We shall derive the minimum error
for discriminating between two ensembles of quantum states from
the no-signalling constraint. The proof is built on a
communication scenario between two parties, Alice and Bob.


\section{Preliminaries}

\begin{figure}
\includegraphics[width=0.32\textwidth]{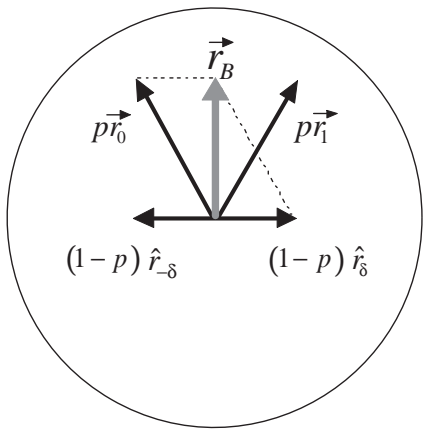}
\caption{There could be infinite number of decompositions for
$\rho(\vec{r}_{B})$, and here we restrict to two decompositions
that can be seen in a cross-section of the Bloch sphere spanned by
$\hat{r}_{ \pm\delta}$ and $\vec{r_{0}}$ or $\vec{r_{1}}$. As is
shown in (\ref{vec}), $\rho(\vec{r}_{B})$ can be a mixture of
$p\vec{r}_{0}$ and $(1-p)\hat{r}_{\delta}$, or $p\vec{r}_{1}$ and
$(1-p)\hat{r}_{-\delta}$, both of which point to the same point in
the Bloch sphere, meaning that they have the same ensemble
average.} \label{fig}
\end{figure}

Before starting the main proof, for clarity we first fix notations
to be used in the communication scenario and then briefly review
some known facts about quantum states, minimum-error
discrimination and ensemble decompositions. We also restrict our
consideration to qubit states, so one can make use of the
representation in which a single qubit state is fully
characterised by its Bloch vector $\vec{v}$: $ \rho(\vec{v}) =
(\openone + \vec{v}\cdot \vec{\sigma})/2$ where $\vec{\sigma} =
(\sigma_{x}, \sigma_{y}, \sigma_{z})$. Pure states have unit Bloch
vectors $\hat{v}$. The communication scenario to be considered
assumes that, sharing entangled states with Bob, Alice measures
her qubits to prepare the ensemble decomposition of Bob's state,
either one of two ensembles, $\rho_{B}^{(0)}$ and
$\rho_{B}^{(1)}$, as follows, \bea
\rho_{B}^{(0)} & = & p\rho_{0} + (1-p)|\delta\rangle \langle \delta|, \nonumber \\
\rho_{B}^{(1)} & = & p\rho_{1} + (1-p)|-\delta\rangle \langle
-\delta|, \label{bobs} \eea where $\rho_{0}$ and $\rho_{1}$ are
two qubit states for which we are to derive the discrimination
bound using the no-signalling constraint, and $|\pm\delta\rangle$
are two orthogonal quantum states such that it holds
$\rho_{B}^{(0)}=\rho_{B}^{(1)}$, see Fig \ref{fig}. It is
convenient to express those quantum states in (\ref{bobs}) in
terms of Bloch vectors as follows, \bea \rho_{B}^{(i)} =
\rho_{B}^{(i)} (\vec{r}_{B}^{(i)}), ~~ \rho_{i} = \rho_{i}
(\vec{r}_{i}), ~~ |\pm\delta\rangle\langle \pm\delta| =
\rho(\hat{r}_{\pm\delta}), \label{parameter} \eea where $i=0,1$.
Then, the relation between quantum states in (\ref{bobs}) can be
translated to the following relations between Bloch vectors \bea
\vec{r}_{B}^{(0)} & = & p \vec{r}_{0} + (1-p) \hat{r}_{\delta},
\nonumber \\
\vec{r}_{B}^{(1)} & = & p \vec{r}_{1} + (1-p) \hat{r}_{-\delta}.
\label{vec}\eea

Recall that $\rho_{0}(\vec{r}_{0})$ and $\rho_{1}(\vec{r}_{1})$ in
(\ref{bobs}) are two quantum states that we wish to discriminate
between. The minimum-error discrimination between two quantum
states has been completely analyzed in Ref. \cite{helstrom}
optimizing over all possible measurement bases. The minimum error,
known as the Helstrom bound, is \bea p_{e}= \frac{1}{2} -
\frac{1}{4}\|\rho_0 - \rho_1\|,\label{hel} \eea assuming that the
\emph{a priori} probability that the quantum state $\rho_{j}$
occurs is equally $1/2$. Here the trace norm for an operator $A$,
denoted by $\|A\|$, is defined as $\| A\| = tr
[\sqrt{A^{\dagger}A}]$. As is shown in (\ref{hel}), the optimal
discrimination between two quantum states only depends on their
(trace-norm) distance separation, and in particular it does not
depend on the dimensions of the states.

We now review properties of \emph{ensembles} of quantum states,
known as mixed states. These are classical mixtures of some pure
quantum states. For instance, each mixed state $\rho_{i}$ (for
$i=0,1$) in (\ref{parameter}) can be expressed as a convex
combination of some other pure states $\{\eta_{j}, |v_{j}^{i}
\rangle \}$, i.e. \bea\rho_{i} = \sum_{j} \eta_{j} |v_{j}^{i}
\rangle \langle v_{j}^{i}|. \label{pure} \eea Note that such a
decomposition is not unique \cite{ghjw}. This means that, in
reality, once an ensemble of quantum states is given, its actual
decomposition cannot be known without knowing how the state was
prepared. One can only deduce ensemble-averaged properties of the
state, which does not depend on any particular decomposition.

Now let us assume that two quantum states in (\ref{bobs}) are the
same. Using the expression in (\ref{vec}), one can see that the
vector of each ensemble of quantum states points to the same point
in the Bloch sphere, see Fig. \ref{fig}. That is, the two quantum
states have different decompositions but have the same ensemble
average. In addition, each $\rho_{j}$, $j=0,1$, is again a mixture
of pure quantum states, so finally quantum states of Bob in
$(\ref{bobs})$ can be expressed as an ensemble of pure quantum
states. Recall that for any mixed state, there exists
higher-dimensional pure quantum state such that partial reduction
of the pure state would be the mixed state itself. In this case,
two-dimensional ancillary systems suffice to purify those quantum
states in (\ref{bobs}). For two different decompositions of the
same state in (\ref{bobs}), the corresponding purifications are
equivalent up to local unitary transformations. This simply
depends on choice of the measurement basis on the ancillary
systems.

If we assume that Alice holds the purification, her measurement
will decide the decomposition of Bob's quantum state. However,
unless Alice announces what basis her measurement was made in, Bob
will never know which decomposition he has. Performing quantum
state tomography will only allow Bob to understand his quantum
state as an ensemble average of other quantum states. This is in
fact what prevents the two parties communicating faster than
light.

\section{State discrimination bound by the no-signalling constraint}

We are ready to prove that the no-signalling constraint implies that
the Helstrom bound in (\ref{hel}) is optimal. Consider that Alice
and Bob are separated in space so that local actions performed by
one cannot affect the other. Suppose that they share copies of
entangled states of $ \C^{2}\otimes \C^{2}$ \bea |\psi\rangle_{AB} =
\sum_{j}\sqrt{\lambda_{j}}|a_{j}\rangle_{A}
|b_{j}\rangle_{B},\label{ab} \eea which is not written in
orthonormal basis of Alice and Bob, so $j$ may exceed the rank of
systems belonging to two parties. As we have mentioned above, Alice
chooses measurement basis so that after the measurement on her
particle, the ensemble decomposition of Bob's state is determined.
In other words, Alice performs a measurement $M_{0}$ or
$M_{1}$(which are general positive-operator-valued-measures), after
which decomposition of Bob's state will be either $\rho_{B}^{(0)}$
or $\rho_{B}^{(1)}$ in (\ref{bobs}) respectively. Alice actually can make use of the measurement $M_{0}$ and $M_{1}$ as an encoding such that a value $j$ is encoded by applying $M_{j}$. Bob then decodes the value by discriminating between two ensembles $\rho_{B}^{(0)}$ and $\rho_{B}^{(1)}$. Two ensembles are the same, i.e. \bea \rho_{B}^{(0)} = \rho_{B}^{(1)} \nonumber \eea and so their Bloch vectors are also the same, i.e.
$\vec{r}_{B}^{(0)} = \vec{r}_{B}^{(1)}$ from (\ref{vec}), and therefore the value $p$ in (\ref{vec}) is uniquely determined, \bea
p=\frac{2}{\|\vec{r}_{0} - \vec{r}_{1}\|+2}. \label{p}\eea


Let us now consider a detector that Bob applies, that works as follows. Once a qubit in a state $\rho$ being measured, the detector gives an outcome $i$ with some probability $P_i (\rho)$ where $i=0,1$, i.e. the detector can be thought of being in a black box from which we only know the input/output list. Therefore, it fulfills that
\begin{equation}
\label{Y} P_0 (\rho)+ P_1 (\rho)=1.
\end{equation}
Note that we do not know all the properties of the detector, e.g. the internal structure, etc., but the
probabilities $P_i (\rho)$ for $i=0,1$. For two states $\rho_{0}$ and $\rho_{1}$, we can assume that in general $P_0 (\rho_0) \geq P_0 (\rho_1)$, by which it also follows that $P_1
(\rho_0) \leq P_1 (\rho_1)$. While Alice not announcing which measurement is applied, Bob can make guess of it by designing his detector to discriminate the two states $\rho_0$ of $\rho_{B}^{(0)}$  and $\rho_1$ of $\rho_{B}^{(1)}$. If the detector discriminates between $\rho_{0}$ and $\rho_{1}$ and gives an output $i$ (either $0$ or $1$), Bob will then conclude that the ensemble is $\rho_{B}^{(i)}$. Note that
Alice applies measurement $M_0$ or $M_1$ with equal probabilities, and then the error rate is
\begin{equation}
\label{Z} e= [P_1 (\rho_0)+ P_0 (\rho_1)]/2.
\end{equation}

Two ensembles in (\ref{bobs}) consist of not only $\rho_{i}$ but also $|\pm\delta \rangle \langle \pm\delta|$. Although the detector being designed to discriminate between the
two states $\rho_0$ and $\rho_1$ in the ensemble, the detector produces some outcomes for $|\delta \rangle
\langle \delta|$ and $|-\delta \rangle \langle -\delta|$ with some
probabilities $P_i (|\delta \rangle \langle \delta|)$ and $P_i
(|-\delta \rangle \langle -\delta|)$ respectively. Assume first that Alice applies the measurement $M_{0}$ so that $\rho_B^{(0)}$ is prepared on the Bob's side. Bob is then with the state $\rho_0$ with probability $p$, from which the value $0$ is found with probability $ P_0(\rho_0)$. Bob is also with the state
$|-\delta \rangle \langle -\delta|$ with probability $1-p$, from which he can guess the value $0$ with probability $P_0(|-\delta \rangle \langle -\delta|)$, a non-negative number. Finally, the overall
probability that detector's output says the value  $0$ when Alice prepares the ensemble $\rho_{B}^{(0)}$, denoted by $D_{0}^{0}$ \footnote{
Let $D_{i}^{j}$ denote the probability that the detector says $i$ when Alice prepares $\rho_{B}^{(j)}$ for $i=0,1$ and $j=0,1$.} and $D_{0}^{0} = p \hspace{1mm} P_0(\rho_0) + (1-p) P_0(|-\delta \rangle \langle -\delta|)$, is greater than or equal to $p \hspace{1mm} P_0(\rho_0)$, i.e.,
\begin{equation}
\label{A} D_0^0 \geq p \hspace{1mm} P_0(\rho_0).
\end{equation}
We also have $D_1^0$, the overall probability that detector's output says the value $1$ when the same ensemble $\rho_{B}^{(0)}$ is prepared. Clearly it holds that
\begin{equation}
\label{B} D_0^0+D_1^0=1.
\end{equation}
Next, assume that Alice applies the measurement $M_{1}$ so that $\rho_{B}^{(1)}$ is prepared. In the same way we have done for the ensemble $\rho_{B}^{(0)}$, the overall probability that the detector says the value $1$, denoted by $D_1^1$, is greater than or equal to $ p \hspace{1mm}
P_1(\rho_1)$, i.e.,
\begin{equation}
\label{C} D_1^1 \geq p \hspace{1mm} P_1(\rho_1).
\end{equation}
It holds again that
\begin{equation}
\label{D} D_0^1+D_1^1=1.
\end{equation}
However, if the following holds
\begin{equation}
\label{E} D_0^0+D_1^1 > 1,
\end{equation}
Bob can discriminate the two ensembles since equations in (\ref{B}), (\ref{D}), and (\ref{E}) tell that
\begin{equation}
\label{F} D_0^0> D_0^1,\hspace{2mm} D_1^0< D_1^1.
\end{equation}
The above (\ref{F}) means that, once $M_{j}$ is applied so that $\rho_{B}^{(j)}$ is prepared, the measurement data show more frequency on the value $j$ than $j+1 (mod ~2)$, through which the two ensemble can be discriminated.
This implies that the no-signalling principle excludes (\ref{E}) and therefore the following holds,
\begin{equation}
\label{G} D_0^0+D_1^1 \leq 1.
\end{equation}
Combining (\ref{A}), (\ref{C}), and (\ref{G}), we have
\begin{equation}
\label{H}  p (P_0(\rho_0)+ P_1(\rho_1)) \leq 1.
\end{equation}
As well, from (\ref{Y}),(\ref{Z}), and (\ref{H}) we obtain
\begin{equation}
\label{I}  e \geq 1- \frac{1}{2p}.
\end{equation}
Finally from (\ref{p}) and (\ref{I}), the following is the discrimination error limited by the non-signalling principle
\begin{equation}
\label{J}  e \geq \frac{1}{2}- \frac{1}{4} \|\vec{r}_{0} -
\vec{r}_{1}\|,
\end{equation}
which is already saturated by the Helstrom bound in (\ref{hel}).

\section{Conclusion}

In conclusion, we relate two no-go theorems - the no-perfect state
discrimination and the no-signalling - by showing that the optimal
discrimination between two quantum states is derived from the
no-signalling condition. It can thus be said that the
impossibility of perfect state discrimination is a consequence of
the no-signalling condition, not only qualitatively, but also
quantitatively as is the case with the no-cloning theorem
\cite{nocloning-gisin}. Furthermore, it was shown in Ref.
\cite{jbae} that the fidelity of optimal quantum cloning converges
asymptotically to the fidelity obtained through optimal state
estimation. Therefore, the quantitative relationship between three
no-go theorems- no-signalling, no-cloning, and no-perfect state
estimation- has been established at the most basic level i.e. two
non-orthogonal quantum states. It would be interesting to investigate whether
optimal values of all figures of merit in state discrimination,
more generally state estimation, of quantum states could be
derived solely from the no-signalling condition. Recently, it is shown that
the optimal measurement for the maximum confidence is also implied by the no-signalling condition
\cite{steph2}. 


\section*{Acknowledgement}
We thank J. Hur, D. Song, and H. Nha for discussions. This work is
supported by the IT R$\&$D program of MKE/IITA (2008-F-035-01), the Korean
Government (MOEHRD) (KRF-2005-003-C00047), and the Korea Science
and Engineering Foundation (R01-2006-000-10354-0).


\end{document}